\documentclass[runningheads]{llncs}

\usepackage[T1]{fontenc}
\usepackage{enumitem}
\usepackage{amsmath}

\usepackage{amssymb}
\usepackage{algorithm2e}
\usepackage{algorithmic}
\usepackage{xcolor}
\usepackage{siunitx}
\usepackage{subcaption}
\usepackage{tabularx}
\usepackage{tcolorbox}
\tcbuselibrary{breakable}
\usepackage{forest}
\usepackage{booktabs}
\usepackage{tikz}
\usetikzlibrary{
    arrows.meta,
    positioning,
    shapes.geometric,
    shapes.misc,
    fit,
    calc,
    backgrounds,
    decorations.pathreplacing
}
\newtcolorbox{featurebox}{
  colback=gray!5,
  colframe=black!40,
  arc=1.5pt,
  boxrule=0.2pt
}

\definecolor{sdggreen}{rgb}{0.0, 0.6, 0.6}

\begin{document}
\title{Feasible Plan Generation with Ambiguity-Boundedness in Cross-Model Query Processing}

\author{Subhasis Dasgupta\inst{1}\orcidID{0000-0002-0754-0515} \and
Amarnath Gupta\inst{1}\orcidID{0000-0003-0897-120X}}

\authorrunning{S. Dasgupta and A. Gupta}

\institute{
University of California San Diego, La Jolla, CA, USA\\
\email{sudasgupta@ucsd.edu, a1gupta@ucsd.edu}
}

%
%
\maketitle              
\begin{abstract}
Natural language (NL) interfaces to databases broaden access to heterogeneous data but often yield many ambiguous intermediate logical plans (ILPs) due to uncertain operator scope and predicate semantics. Many candidates are infeasible because of type mismatches, missing bindings, or engine-specific constraints. We address this challenge with \emph{feasibility constraints} for detecting local inconsistencies and introduce the \emph{Packed Plan Forest (PPF)}—a polynomially bounded structure that compactly encodes all feasible ILPs while pruning infeasible ones early. Extending packed parse forest ideas to multi-model settings, PPF supports efficient feasibility analysis through annotated operators. Formal results show polynomial size under bounded arity and annotation vocabularies, and experiments confirm that PPFs capture exponentially many ILPs with minimal overhead, establishing a scalable foundation for NL-to-DB query planning across heterogeneous systems.
\end{abstract}
\section{Introduction}
\label{sec:intro}
Modern analytical infrastructures are shifting toward multi-model, heterogeneously managed data systems, also known as data lakes (e.g., Apache Hudi, Lakehouse, DuckLak), where relational, graph, vector, and spatial layers coexist and interoperate through a unified query interface \cite{barret2023user,yuan2023effective}. These systems power domains such as data , supply-chain traceability, health informatics, and scientific knowledge networks and many more, where users express semantic queries in natural language over multi-engine backends.\\ Traditional NL2DB pipelines \cite{dar2019frameworks} typically assume a homogeneous data setting governed by a single logical model, where ambiguity resolution and plan generation are treated as separate stages. In contrast, heterogeneous data ecosystems introduce a deeper coupling between semantic interpretation and execution feasibility. However, a syntactically valid interpretation may prove infeasible in execution due to missing entity bindings, coordinate reference inconsistencies, or absent vector embeddings across heterogeneous engines. Such interdependence demands joint reasoning over semantics, feasibility, and cost.
\\[4pt]
\noindent\textbf{Running Example:} Imagine a Sustainable Supply Chain Intelligence (SSCI) system that traces the lifelines of global production—tracking suppliers, parts, projects, and the web of sustainability documents that bind them—across a constellation of semantically interconnected engines:
\begin{featurebox}
\begin{verbatim}
Relational Data:
    Suppliers(sid, name, address, docid)
    Parts(pid, name, category)
    Projects(projid, name, address)
Property Graph:
    Supplier, Part, Project nodes with edges
    Supplies(Supplier→Part), UsedIn(Part→Project)
Vector Store:
    Documents(docid, text, embedding)
Geospatial:
    AdminBoundaries(geom, country, crs), 
    geocoding/CRS transforms
\end{verbatim}
\end{featurebox}
\noindent
These components span: (i) a relational engine (e.g., PostgreSQL) for base records, (ii) a graph engine (e.g., Neo4j) for supply relationships, (iii) a vector store (e.g., Qdrant) for sustainability documents, and (iv) spatial services (e.g., PostGIS) for geometries and coordinate transforms.\\
With a natural language interface, a user may pose:  
\textit{“Find suppliers of parts used in projects in Germany with documents discussing sustainability.”}  
NL-to-DB systems \cite{LiJagadish2014VLDB,Yaghmazadeh2017POPL,Affolter2019VLDBJ,Guo2019ACL} typically generate multiple candidate logical forms (intermediate logical plans, ILPs), which are then ranked or pruned. In a heterogeneous setting, however, many of these candidate ILPs are infeasible due to type inconsistencies, unbound entities, coordinate reference misalignments, or missing vector embeddings.
Importantly, semantic ambiguity and planning variation are intertwined operator placement and annotation choices jointly determine feasibility. Thus, it becomes necessary to eliminate infeasible branches at an early stage while still retaining all viable interpretations.
\begin{featurebox}\it
    How can we compactly represent this large space of candidate query plans, enforce heterogeneous feasibility constraints early, and efficiently decide whether any feasible execution plan exists?
\end{featurebox}
\noindent
We address this challenge with the \emph{Packed Plan Forest (PPF)}, a compact representation inspired by parse forests in NLP and AND–OR DAGs in query optimization. The PPF collapses exponentially many ILPs into polynomial space by sharing common subplans across relational, graph, spatial, and vector operators. Feasibility is defined via local constraints on types, bindings, CRS alignment, engine placement, and uncertainty. A bottom-up labeling algorithm checks feasibility or diagnoses infeasibility in polynomial time. In summary, this work makes the following key contributions:
\begin{enumerate}[ leftmargin=*]
\item An annotated-operator framework for heterogeneous feasibility constraints (types, bindings, CRS, engine placement, uncertainty) across relational, graph, vector, and spatial models.
\item An algorithmic framework that \emph{bridges natural language ambiguity} and systematically generates candidate intermediate logical plans across heterogeneous data models, including relational, graph, vector, and spatial systems.
\item The \emph{Packed Plan Forest (PPF)}: a polynomial-space structure that compactly encodes all feasible ILPs via shared subplans and early infeasibility pruning.
\item Formal polynomial bounds on the PPF size and feasibility-checking complexity, with experimental evidence of near-linear scaling and exponential compression.
\end{enumerate}
We next survey related work on feasibility reasoning in polystores, compact plan representations, and NL interfaces.
\section{Related Work}
\label{sec:related}
The problem of joint reasoning over ambiguity and feasibility in heterogeneous query planning spans three research areas.

\smallskip\noindent\textbf{Feasibility in polystores.}
Early work (Information Manifold, TSIMMIS~\cite{levy1996answering,halevy2001answering}) formalized the answerability of the query under source bindings as logical dependencies. Benedikt et al.~\cite{benedikt2017query} treated execution as constraint satisfaction over functional dependencies. BigDAWG~\cite{stonebraker2018bigdawg}, AWESOME~\cite{dasgupta2016analytics}, and CloudMdsQL\cite{kolev2016cloudmdsql} extend this to multi-model settings but lack formal infeasibility witnesses. Our approach makes feasibility declarative: local annotations capture types, bindings, CRS alignment, and engine placement, and failures yield structured certificates.
\smallskip\noindent\textbf{Compact plan representations.}
Volcano/Cascades~\cite{graefe1993volcano} use memo structures and AND-OR DAGs to share equivalent subplans. Packed parse forests~\cite{tomita1987efficient} adapt this to NLP ambiguity. PPF extends these to multi-model settings: annotated operator nodes and feasibility-constraint edges give PPF a dual role as both a memoization structure and a symbolic encoding of the constraint space (Section~\ref{sec:ILP}).
\smallskip\noindent\textbf{NL interfaces and ambiguity.}
Handling multiple interpretations of user intent has been a persistent theme in NLIDB research, which enumerates alternative logical forms and uses interaction to resolve scope~\cite{li2014constructing}. Probabilistic and
neural text-to-SQL systems~\cite{zhong2017seq2sql,yaghmazadeh2017sqlizer} replace rule-based enumeration with ranking over candidate SQL programs. However, they target homogeneous schemas and treat feasibility as type consistency. Our work extends the ambiguity to the operator-model and engine-placement choices, making feasibility integral to plan generation.

\begin{table}[t]
\centering
\scriptsize
\caption{Operator families for Intermediate Logical Plans (ILPs). 
Notation: $R$=rel, $V$=nodes, $E$=edges, $P$=paths, $D$=docs, $G$=geom.}
\renewcommand{\arraystretch}{0.95}
\setlength{\tabcolsep}{1.8pt}
\begin{tabularx}{\columnwidth}{|l|l|X|}
\hline
\textbf{Op} & \textbf{Sig.} & \textbf{Brief Description} \\
\hline
\multicolumn{3}{|c|}{\textbf{Relational}}\\
\hline
$\sigma_{\theta}$ & $R\!\to\!R$ & Filter tuples by $\theta$.\\
$\pi_A$ & $R\!\to\!R$ & Project attrs $A$.\\
$R_1\!\bowtie_{\theta}\!R_2$ & $R^2\!\to\!R$ & Join on $\theta$.\\
$\gamma_{A,f}$ & $R\!\to\!R$ & Group by $A$, agg. $f$.\\
\hline
\multicolumn{3}{|c|}{\textbf{Graph}}\\
\hline
$\tau_E$ & $V\!\to\!V$ & Traverse via edges $E$.\\
$\mu_{\pi}$ & $V\!\to\!P$ & Match pattern $\pi$.\\
\hline
\multicolumn{3}{|c|}{\textbf{Text / Vector}}\\
\hline
$\kappa_T$ & $D\!\to\!D$ & Keyword filter $T$.\\
$\phi_E$ & $D\!\to\!D$ & Full-text / ranked expr.\\
$\nu^k_{\vec{q}}$ & $D\!\to\!D$ & Top-$k$ similar to $\vec{q}$.\\
$\varsigma_{\vec{q}}$ & $D^2\!\to\!R$ & Pair docs by sim. thresh.\\
\hline
\multicolumn{3}{|c|}{\textbf{Spatial / GIS}}\\
\hline
$\sigma^S_{\phi}$ & $G\!\to\!G$ & Select geom. by $\phi$.\\
$G_1\!\bowtie^S_{\phi}\!G_2$ & $G^2\!\to\!G$ & Spatial join.\\
$\rho^{CRS}_{c\to c'}$ & $G\!\to\!G$ & Reproject CRS.\\
$\kappa^S_{nn}$ & $G\!\to\!G$ & $k$-NN query.\\
\hline
\multicolumn{3}{|c|}{\textbf{Semantic / LLM}}\\
\hline
$\eta_{\text{type}}$ & $D\!\to\!R$ & Extract entities.\\
$\rho_{\text{type}}$ & $D\!\to\!E$ & Extract relations.\\
$\chi_{\text{label}}$ & $D\!\to\!R$ & Classify docs.\\
$\varsigma^{LLM}_{task}$ & $D\!\to\!D$ & LLM enrich (embed/sum).\\
\hline
\multicolumn{3}{|c|}{\textbf{Cross-Model}}\\
\hline
$\xi_{R\to V}$ & $R\!\to\!V$ & Tuples→graph.\\
$\xi_{V\to R}$ & $(V,E)\!\to\!R$ & Graph→table.\\
$\xi_{R\to D}$ & $R\!\to\!D$ & Add text/embeds.\\
$\xi_{D\to R}$ & $D\!\to\!R$ & Sim. join / IDs.\\
$\xi_{R\to G}$ & $R\!\to\!G$ & Geocode addr.\\
$\xi_{G\to R}$ & $G\!\to\!R$ & Extract coords.\\
\hline
\end{tabularx}
\label{tab:operators}
\end{table}

\section{Intermediate Logical Plans}
\label{sec:ILP}
Classical logical plans built on a single algebra are inadequate for our setting, while Natural language (NL) queries over heterogeneous data involve operators from multiple models. Their feasibility depends on constraints not expressible in relational signatures alone. We therefore introduce the \emph{Intermediate Logical Plan (ILP)} as a unifying representation. ILPs generalize logical plans in two ways: (i) they draw operators from relational, graph vector, spatial, and semantic domains, linked by explicit cross-model casts, and (ii) they carry \emph{annotations} recording typing, bindings, coordinate systems, placement, uncertainty, and semantic tags. ILPs form the algebraic objects over which we reason about feasibility (Section~\ref{sec:feasible}) and construct compact representations (Section~\ref{sec:ppf}).
\subsection{Operator Algebra}
\label{sec:algebra}
An ILP is an expression over a family of typed operators. Each operator has a signature
$o : (I_1,\ldots,I_m) \to (O_1,\ldots,O_n)$,
where inputs $I_j$ and outputs $O_\ell$ are typed objects such as, relations, graph nodes/edges, paths, documents, embeddings or geometries. ILPs are composed into directed acyclic graphs (DAGs) where nodes are operators, and edges
represent typed data flow.
\begin{featurebox}
  \textbf{Relational:} $\sigma,\pi,\bowtie,\gamma$ for selection, projection, joins, aggregation.\\
  \textbf{Graph:} $\tau$ for traversals and $\mu$ for path/pattern matching.\\
  \textbf{Vector/Text:} $\kappa,\phi,\nu,\varsigma$ for keyword search,
  full-text, $k$-NN embedding search, and similarity joins.\\
  \textbf{Spatial/GIS:} $\sigma^S,\bowtie^S,\rho^{CRS}$ for spatial filters, joins, and CRS reprojections.\\
  \textbf{Cross-model:} $\xi$ operators (e.g., $\xi_{D \to R}$,
  $\xi_{R \to G}$) provide bridges across domains.\\
  \textbf{Semantic/LLM:} optional enrichment operators
  ($\eta,\rho,\chi,\varsigma^{LLM}$) extend the algebra without changing core feasibility analysis.   
\end{featurebox}  
\noindent
Cross-model operators are especially critical. For example,
$\xi_{R \to V}$ casts relational tuples into graph nodes, while $\xi_{R \to G}$
geocodes addresses into geometries. Without such operators, ILPs remain
confined to a single model, and feasibility analysis would be trivial.

\medskip
\noindent\textbf{Example 1.}
\textit{``List projects that use parts supplied by companies with documents similar to the sustainability report.''}
\[
\begin{array}{c}
\pi_{\text{projid},\, \text{name}} \\
\Big| \\
\bowtie \\
/ \quad \backslash \\
\tau_{\text{UsedIn}} \quad Projects \\
\Big| \\
\tau_{\text{Supplies}} \\
\Big| \\
\xi_{D \rightarrow R} \\
\Big| \\
\nu^{k}_{\vec{q}_{\text{sust}}}(Docs)
\end{array}
\]
\noindent
The ILP retrieves sustainability-related documents with 
$\nu^k_{\vec{q_{sust}}}$, maps them back to suppliers using 
$\xi_{D \to R}$, traverses the supply chain through 
$\tau_{\text{Supplies}}$ and $\tau_{\text{UsedIn}}$, and finally 
joins with the \texttt{Projects} relation. Annotations drive 
feasibility at each step: schema metadata declares that 
$\xi_{D \to R}$ must resolve \texttt{docid} to \texttt{sid}, the 
system catalog places $\nu^k$ in Qdrant and $\tau$ operators in 
Neo4j, and the operator template propagates the 
$\epsilon$-approximate semantics of $\nu^k$ so that joins expecting 
deterministic identifiers remain valid. If the catalog mapping 
\texttt{docid} $\mapsto$ \texttt{sid} is missing, or if Qdrant does not 
support the required embedding index, infeasibility is reported with 
a certificate. Otherwise, the plan is retained for further packing 
and optimization.

\medskip
\noindent\textbf{Example 2.}
\textit{``Which suppliers connected to projects in Europe have documents similar to the EU sustainability guidelines?''}

\begin{forest}
for tree={
    draw,
    rounded corners,
    align=center,
    parent anchor=south,
    child anchor=north,
    l sep=12pt,
    s sep=12pt,
    edge={->}
}
[$\pi_{\text{sid},\,\text{name}}$
    [$\bowtie$
        [$\bowtie$
            [$\bowtie$
                [$\tau_{\text{Supplies}}(Suppliers)$]
                [$\tau_{\text{UsedIn}}(Parts)$]
            ]
            [$\sigma^{S}_{\text{within(Europe)}}$
                [$\xi_{R \rightarrow G}(Project)$]
            ]
        ]
        [$\xi_{D \rightarrow R}$
            [$\nu^{k}_{\vec{q}_{\text{EU}}}(Docs)$]
        ]
    ]
]
\end{forest}\\[4pt]
\noindent
The ILP first retrieves EU-related documents via $\nu^k_{\vec{q_{EU}}}$,
casts matches to supplier/project IDs with $\xi_{D \to R}$, traverses the
supply chain in the graph ($\tau_{\text{Supplies}}$, $\tau_{\text{UsedIn}}$),
and filters projects spatially after geocoding ($\xi_{R \to G}$ followed by
$\sigma^S_{\text{within}}$). Feasibility follows from the three annotation
sources: (a). schema metadata expose and repair CRS misalignment (details in section \ref{sec:annotations})  (b). the system catalog binds each operator to a supporting engine, and (c). the operator template ensures
that $\epsilon$-approximate semantics from $\nu^k$ are propagated safely
through $\xi_{D \to R}$ and downstream joins. If any offaile cTheks fails,
the plan is rejected with a localized witness; otherwise, the plan remains in
the candidate set for packing (Section~\ref{sec:ppf}) and cost-based selection.

\medskip
\noindent\textbf{Example 3.}
\textit{``List suppliers that provide components to projects whose funding
documents are semantically similar to EU policy guidelines, and whose project
timelines overlap with FY2024.''}

\begin{forest}
for tree={
    draw,
    rounded corners,
    align=center,
    parent anchor=south,
    child anchor=north,
    l sep=12pt,
    s sep=12pt,
    edge={->}
}
[$\pi_{\text{sid},\,\text{name}}$
    [$\bowtie$
        [$\bowtie$
            [$\bowtie$
                [$\xi_{D \rightarrow R}$
                    [$\nu^{k}_{\vec{q}_{\text{EU}}}(FundingDocs)$]
                ]
                [$\tau_{\text{Supplies}}(Suppliers)$]
            ]
            [$\tau_{\text{UsedIn}}(Parts)$]
        ]
        [$\sigma^{T}_{\text{overlap(FY2024)}}$
            [$\xi_{R \rightarrow T}(Projects)$]
        ]
    ]
]
\end{forest}\\
\noindent
This ILP combines vector similarity, graph traversals, and temporal filters. 
The operator $\nu^k_{\vec{q_{EU}}}$ retrieves funding documents aligned with 
EU policy embeddings; annotations here come from operator templates (error 
bounds and embedding domain) and catalog entries (engine = Qdrant). The cross-model 
operator $\xi_{D \to R}$ maps document identifiers back to suppliers, requiring 
schema metadata that defines the \texttt{docid} $\mapsto$ \texttt{sid} mapping. 
The graph traversals $\tau_{\text{Supp}}$ and $\tau_{\text{UsedIn}}$ 
consume suppliers and parts, annotated with node and edge types from the schema 
and placement in the Neo4j engine. Finally, $\xi_{R \to T}(Proj)$ produces 
temporal intervals, with schema metadata providing date domains, and the operator 
$\sigma^T_{\text{overlap(FY2024)}}$ applies a fiscal-year filter whose semantics 
are drawn from the operator template. \\
Feasibility depends on consistent alignment 
across these domains: (i) whether the embeddings for funding documents are tagged 
as compatible with EU policy guidelines, (ii) whether temporal annotations 
calendar vs.\ fiscal align or require normalization, and (iii) whether the catalog 
confirms that each operator is supported on its designated engine. If any of these 
checks fail, infeasibility certificates isolate the violation (e.g., embedding 
domain mismatch or fiscal vs.\ calendar misalignment); otherwise, the plan is 
retained as a feasible candidate.
\subsection{Annotations}
\label{sec:annotations}
Besides the operators, metadata annotations of the nodes are a key factor in determining feasibility.  We therefore enrich each node with an annotation vector $\alpha(o)$ capturing. Hence, the operator metadata profile is as follows: 
\begin{featurebox}
\textbf{Type profile:} schema, graph label, geometry type \\
\textbf{Binding status:} query inputs bound/unbound \\
\textbf{Reference system:} CRS, temporal granularity, units\\
\textbf{Placement:} Postgres, Neo4j, Qdrant, PostGIS\\
\textbf{Uncertainty:} deterministic or $\epsilon$-bounded\\
\textbf{Semantic tags:} embedding/domain annotations
\end{featurebox}
\noindent
Annotations are populated from three sources: schema metadata (types, CRS),
system catalogs (engine placement), and operator templates (uncertainty). Some
semantic tags may be derived by lightweight enrichment or user hints. Consider Example 2 where  
annotations are populated from three distinct sources:\\[4pt]
\noindent
\textbf{Schema metadata:}
The operator $\xi_{R \to G}(Proj)$ produces geometries
$[geom:Point,crs=EPSG:4326]$ , while the Europe polygon is stored as $[geom:Point,crs=EPSG:3857]$ A CRS misalignment is detected directly from schema annotations, requiring the insertion of a reprojection operator : $\Large\rho^{\mathrm{CRS}}_{4326 \to 3857}$.\\[4pt]
\textbf{System catalogs:}
The join operator $\bowtie$ and the spatial filter $\sigma^S_{\text{within}}$ must be mapped to execution engines that support them. The catalog records that $\sigma^S$ is admissible only in PostGIS, while $\tau_{\text{Supplies}}$ and $\tau_{\text{UsedIn}}$ require Neo4j.
This ensures distribution constraints are enforced during planning.\\[4pt]
\textbf{Operator templates:}
The vector similarity operator $\nu^k_{\vec{q_{EU}}}$ is annotated with an $\epsilon$-approximate uncertainty profile by template. The downstream cross-model operator $\xi_{D \to R}$ checks that approximate identifiers can be propagated safely, preserving error bounds in subsequent joins.\\[4pt]
\noindent
Some operators expose continuous parameters (e.g., vector similarity thresholds, spatial distances, or temporal overlap tolerances). In such cases, annotation domains can be discretized into a finite representative set following established practices in query optimizers, spatial indexing, and approximate nearest-neighbor search. Thresholds may be rounded into system-defined bins, distance predicates mapped to index-specific radii, and embedding similarity cutoffs approximated using quantized score intervals. This abstraction preserves the finiteness assumption required for polynomial feasibility checking, while aligning with real-world execution semantics.
\subsection{Candidate ILPs and Attachments}
\label{sec:candidate}
A natural language query seldom maps to a single ILP. Instead, parsers produce
\emph{families} of candidates differing in where predicates attach, how they
are interpreted, and which operators are chosen.\\[4pt]
\noindent\textbf{Attachment:}
We define \emph{attachment} as the locus in the ILP where a predicate from the
NL query is bound. For instance, ``documents similar to the sustainability
report'' may attach to the \emph{Supplier} relation (supplier-level documents)
or to \emph{Projects} (project-level documents). Each attachment yields a
different ILP structure.\\[4pt]
\noindent\textbf{Sources of multiplicity:}
Source multiplicity arises from the scope and interpretation of source predicates under semantic mappings. Such mappings may involve value transformations (e.g., normalization or scaling), reinterpretation of predicate semantics, or alternative operator realizations. Consequently, multiplicity manifests along several dimensions: (i) \emph{attachment scope}, where data may be associated with different entities (e.g., supplier- vs.\ project-level documents); (ii) \emph{predicate interpretation}, where a condition such as “in Europe” may be realized via a spatial join, a country-level filter, or a nearest-neighbor approximation; and (iii) \emph{operator alternatives}, where semantic intent (e.g., “sustainability”) may be evaluated using keyword-based retrieval or vector similarity. These dimensions collectively expand the space of valid execution plans.
\noindent
The ILP algebra and annotations jointly enable us to represent, compare, and
filter candidate plans. Ambiguity arises from attachment, predicate semantics,
and operator choice. Feasibility analysis (next section) builds directly on
these annotations.

\subsection{Infeasible Plans and Certificates}
\label{sec:feasible}
Candidate ILPs generated from NL queries often include plans that cannot be
executed because one or more constraint families from Section~4.2 are violated.
We model infeasibility as a constraint satisfaction problem (CSP) over operator
annotations, and we provide \emph{infeasibility certificates}\footnote{We use
the term certificate in the complexity-theoretic sense of a verifiable witness:
a minimal explanation of infeasibility that can be checked in polynomial time.
In the database literature, analogous notions are often referred to as
``diagnostics'' or ``witnesses.''} as verifiable explanations of failure.\\[4pt]
\noindent\textbf{Witnesses:}
We use the term \emph{witness} to denote the local evidence trace $\pi$
extracted from operator annotations that shows why a particular constraint
failed (e.g., CRS mismatch, type incompatibility, or missing binding). A
witness is the atomic unit of evidence, typically expressed as a tuple of
conflicting annotation values. Witnesses are attached incrementally during plan
construction or pruning.\\[4pt]
\noindent\textbf{Certificates:}
An \emph{infeasibility certificate} aggregates witnesses into a structured
tuple:
$
\mathcal{C} = (P', o, C, \pi),
$
where $P' \subseteq P$ is the minimal subplan exhibiting the violation, $o$ is
the operator at which infeasibility manifests, $C \in \mathcal{F}$ is the
violated constraint family, and $\pi$ is the witness. Minimality means that no
proper subplan of $P'$ is infeasible with respect to $C$. Thus, certificates
formalize infeasibility by pairing a violating operator and constraint with its
local evidence trace.\\
Now let's see what this means for the examples we looked at earlier:\\[4pt]
 \textbf{Type incompatibility.} In Example~1 (sustainability report),
  $\xi_{D \to R}$ expects a mapping $\texttt{docid} \mapsto \texttt{sid}$. If
  the catalog lacks this mapping, the witness is
  $\pi = \{\texttt{docid} \not\mapsto \texttt{sid}\}$. The certificate is\\
  $(\{Supplier,Documents,\xi_{D \to R}\},  \xi_{D \to R}, \texttt{TYPE}, \pi)$.\\[4pt]
  \textbf{Semantic misalignment.} In Example~2 (EU guidelines), supplier
  nodes in the graph are typed \texttt{Company}, while the edge pattern expects
  \texttt{Organization}. The witness is
  $\pi = \{\texttt{Company} \not\subseteq \texttt{Organization}\}$. The
  certificate is \\$(\{Suppliers,\tau_{\text{Supplies}}\}, \tau_{\text{Supplies}},
  \texttt{ALIGN}, \pi)$.\\
  \textbf{Binding failure.} In Example~2, if
  $\nu^k_{\vec{q_{EU}}}$ is unbound, the witness is \\
  $\pi = \{\vec{q_{EU}} \text{ missing}\}$, yielding the certificate
  $(\nu^k_{\vec{q_{EU}}}, \nu^k_{\vec{q_{EU}}}, \texttt{BOUND}, \pi)$.\\
  \textbf{Composite failure.} In Example~3 (funding timelines), two
  independent witnesses arise:
  $\pi_1 = \{\texttt{embedding domains mismatch}\}$ and
  $\pi_2 = \{\texttt{FiscalQ} \neq \texttt{CalendarW}\}$. These form two
  certificates, one for semantic tags and one for temporal alignment.\\
\noindent\textbf{Minimality and locality.}
Certificates are localized: they identify the smallest subplan and operator
responsible, and the witness provides the exact evidence trace. This supports
incremental pruning: during plan generation, as soon as a partial ILP produces
a witness, that branch can be pruned. Minimality ensures non-redundancy:
removing any operator from $P'$ eliminates the violation.\\
\noindent\textbf{Complexity.}
Since constraints are local predicates over operator annotations and their
adjacencies, infeasibility checking is polynomial in $|O|+|E|$. Certificate
generation scales linearly with the number of operators, because witnesses are
recorded as constant-size evidence traces. This tractability underpins the
packed representation of Section~5.\\
\noindent\textbf{Role in planning.}
Certificates, built from witnesses, serve two roles. First, they prune
infeasible ILPs early, avoiding wasted optimization effort. Second, they provide
diagnostic feedback. For Example~2, the system may report:
\emph{``Traversal infeasible: Supplier nodes typed as Company, but edge
WorksOn expects Organization''}. Unlike traditional query optimizers that
reject invalid queries silently, certificates provide verifiable and
human-readable explanations of why a plan cannot execute.
\section{Plan Generation and Validation: Packed Plan Forest Algorithm \& Feasibility Checking Algorithm }
\label{sec:ppf}
The previous section established how feasibility can be determined locally via annotations on operators. However, natural language queries usually generate not a single ILP but a combinatorially large set of candidate ILPs, often growing exponentially with the length of the query. As a result, listing each ILP one by one is impractical. The \emph{Packed Plan Forest (PPF)} is our compact representation of this candidate space. It encodes all feasible ILPs in polynomial space by sharing equivalent subplans and pruning infeasible ones early.
\\[4pt]
Consider a collection of candidate ILPs $\{P_1,\dots,P_m\}$. A straightforward encoding would list each plan as its own DAG. When each of the $n$ predicates in the query has $k$ possible operator choices, the total number of plans is $m=O(k^n)$. The PPF prevents this combinatorial explosion by merging equivalent subplans: any sub-DAGs that use the same operator symbols and have compatible annotations are represented only once in the PPF and are shared among all derivations.\\[4pt]
\noindent
The PPF combines ideas from memoization structures in query optimizers and packed parse forests in NLP. Like an AND--OR DAG, it represents multiple derivations compactly by factoring common subexpressions. Like a parse forest, it stores multiple syntactic alternatives in shared nodes. Our extension is to incorporate annotated operators and feasibility constraints across heterogeneous models. This means that merging is annotation-sensitive: operators are shared only if their annotation vectors (types, CRS, uncertainty, engine placement, semantic tags) are jointly feasible.
\begin{definition}[Packed Plan Forest]
Given a query $Q$, the \emph{Packed Plan Forest (PPF)} is a triple 
$\mathcal{P} = (N,E,\equiv)$, \\ where $N$ is a set of nodes, each corresponding 
to an operator $o$ with an equivalence class of annotations $\alpha(o)$ that are 
jointly feasible; $E \subseteq N \times N$ is a set of directed edges such that 
$(n_1,n_2) \in E$ iff the output of $n_1$ can feed the input of $n_2$ and all 
constraints in Section~\ref{sec:annotations} are satisfied; and $\equiv$ is an 
equivalence relation that identifies nodes representing the same operator 
semantics and compatible annotations, regardless of upstream derivations.
\end{definition}
\subsection{Construction Algorithm}
The PPF is built incrementally from the set of candidate ILPs or from an incremental NL parse. Algorithm~\ref{alg:ppf-build} outlines the construction. Construction of the \emph{Packed Plan Forest (PPF)} from a set of candidate Intermediate Logical Plans (ILPs). The algorithm iteratively computes annotation vectors, applies local feasibility checks, and merges equivalent annotated operators via hash-consing to build a compact AND–OR representation. 
Infeasible branches are pruned early, and the resulting forest $\mathcal{P} = (N, E, \equiv)$ encodes all feasible ILPs in polynomial space. The procedure uses the constraint families of Section~\ref{sec:annotations} as an oracle for early pruning: infeasible operators never enter the forest. The lookup-or-create routine ensures that semantically equivalent subplans are shared, but only if their annotations are compatible. For instance, two vector similarity nodes using different embedding domains (\texttt{EUguidelines} vs.\texttt{FundingDocs}) remain distinct, while two variants differing only in placement (e.g., both supported by Qdrant) can merge.
\begin{algorithm}[t]
\caption{PPF construction from candidate ILPs.}
\label{alg:ppf-build}
\scriptsize
\begin{algorithmic}[1]
\REQUIRE $\{P_1,\dots,P_m\}$
\ENSURE $\mathcal{P}=(N,E,\equiv)$
\STATE $N,E \gets \emptyset$
\FOR{each plan $P_i$}
  \FOR{each operator $o \in P_i$}
    \STATE $\alpha \gets \alpha(o)$
    \IF{feasible$(o,\alpha)$}
      \STATE $n \gets$ lookup-or-create$(o,\alpha,N,\equiv)$
      \STATE add $n$ to $N$ and connect to compatible predecessors in $E$
    \ELSE
      \STATE prune branch and record witness
    \ENDIF
  \ENDFOR
\ENDFOR
\RETURN $(N,E,\equiv)$
\end{algorithmic}
\end{algorithm}
\noindent\textbf{Complexity analysis:}
Let $Q$ be a query with $n$ predicates, each yielding at most $k$ operator candidates. Let $|\mathcal{A}|$ denote the size of the annotation vocabulary (types, CRS codes, semantic tags, etc.). Algorithm~\ref{alg:ppf-build} processes each candidate operator once, computing its annotation vector and performing feasibility checks. Since feasibility constraints are local predicates, each check runs in $O(1)$ time relative to the size of the annotation vector. Node lookup-or-create is performed via hash-consing and can be implemented in $O(1)$ expected time.\\[4pt]
Thus the cost of PPF construction is $O(n \cdot k \cdot |\mathcal{A}|)$. For generating nodes, plus $O(n \cdot k \cdot d \cdot |\mathcal{A}|^2)$, and for connecting edges, where $d$ is the maximum operator arity. Both terms are polynomial in $n$ and $|\mathcal{A}|$, even though the PPF encodes up to $O(k^n)$ distinct ILPs. This formalizes the claim that infeasibility detection and plan packing are tractable.\\[5pt]
\textbf{Polynomial Boundedness:Size of PPF:} Let $Q$ be a query with $n$ predicates. Suppose each predicate admits at most
$k$ operator choices, each operator has arity at most $d$, and the annotation
vocabulary $\mathcal{A}$ has size $|\mathcal{A}|$. Then the size of the PPF is
bounded by : 
$
|N| \leq O(n \cdot k \cdot |\mathcal{A}|), \quad
|E| \leq O(n \cdot k \cdot d \cdot |\mathcal{A}|^2).
$

\begin{algorithm}[t]
\caption{PPF feasibility labeling.}
\label{alg:feasibility}
\scriptsize
\begin{algorithmic}[1]
\REQUIRE $\mathcal{P}=(N,E,\equiv)$, constraints $\mathcal{F}$
\ENSURE $L(n)$ for all $n \in N$
\STATE $L(n) \gets \Lambda(n)\;\forall n$
\REPEAT
  \FOR{$n \in N$ (rev. topo)}
    \STATE $L(n) \gets \{\lambda \in L(n)\mid 
    \forall (m,n)\!\in\!E,\;\exists \mu\!\in\!L(m): 
    \mathcal{F}(\mu,\lambda)\}$
  \ENDFOR
\UNTIL{fixed point}
\RETURN $\{L(n)\}$
\end{algorithmic}
\end{algorithm}

\noindent
\textbf{Feasibility Checking Algorithm (over PPF):}
\label{sec:feasibility-ppf}
The PPF provides a compact representation of all possible ILP formulations. Feasibility can be determined directly on this structure using a bottom-up labeling procedure that operates in polynomial time and, when no valid plan is present, yields formal certificates of infeasibility. \\
\noindent
\textbf{Labeling Algorithm:}
\label{sec:labeling}
Each node $n \in N$ receives a \emph{feasibility label set} $L(n) \subseteq \Lambda(n)$ of annotation configurations for which $n$ can participate in a feasible ILP. Algorithm~\ref{alg:feasibility} propagates these labels bottom-up, pruning configurations inconsistent with any predecessor; sink nodes with $L(n)=\emptyset$ certify infeasibility.\\
\noindent
\textbf{Polynomial feasibility checking : } Let $\mathcal{P}=(N,E,\equiv)$ be a PPF with $|N|$ nodes, maximum operator
arity $d$, and per-node annotation domain size $K$. Then Algorithm~\ref{alg:feasibility} decides feasibility in time $O(|N| \cdot K^d)$.
This is crucial for determining plan space, even though the PPF can represent an exponential number of ILPs, feasibility can still be checked efficiently. Upon termination, every sink node maintains a nonempty label set; consequently, the PPF represents only consistent ILPs, while infeasibility witnesses are associated with discarded annotations.\\
\noindent\textbf{Infeasibility Certificates:} Whenever a label $\lambda \in L(n)$ is pruned, the algorithm records a witness explanations (e.g., ``Embedding domain mismatch'', ``Temporal granularity mismatch'') bridging structural feasibility and diagnostic feedback. Tractability holds under bounded arity, finite annotation domains, and local-only constraints; extending to global constraints requires SAT/SMT reasoning.
\begin{table}[t]
\centering
\scriptsize
\caption{Packed Plan Forest Evaluation.
\textbf{Abbrev.:} \#P = candidate plans; 
N = avg. nodes per plan; 
UniqA = unique nodes (no pruning); 
UniqF = unique feasible nodes; 
PkA = packing ratio (pre-prune); 
PkF = packing ratio (post-prune); 
PrU = pruned unique nodes; 
$t_{ms}$ = runtime (ms) with pruning; 
MemK = peak memory (KB).
}
\renewcommand{\arraystretch}{0.9}
\setlength{\tabcolsep}{3pt}

\resizebox{\columnwidth}{!}{%
\begin{tabular}{lrrrrrrrrr}
\toprule
Scen & \#P & N & UniqA & UniqF & PkA & PkF & PrU & $t_{ms}$ & MemK \\
\midrule
S1 &100&25&2290&1690&0.92&0.68&600&27.5&6301\\
S2 &100&25&2359&1749&0.94&0.70&610&43.2&6409\\
S3 &100&25&2208&1656&0.88&0.66&552&24.2&6310\\
S4 &100&25&2400&693&0.96&0.28&1707&21.3&2708\\
S5 &100&25&2389&1755&0.96&0.70&634&25.8&6366\\
S6 &1000&50&25882&16600&0.52&0.33&9282&808.5&84961\\
\bottomrule
\end{tabular}}
\label{tab:ppf-eval}
\end{table}

\begin{figure}[t]
\centering
\scriptsize

\begin{minipage}{0.48\columnwidth}
\centering
\includegraphics[width=\linewidth]{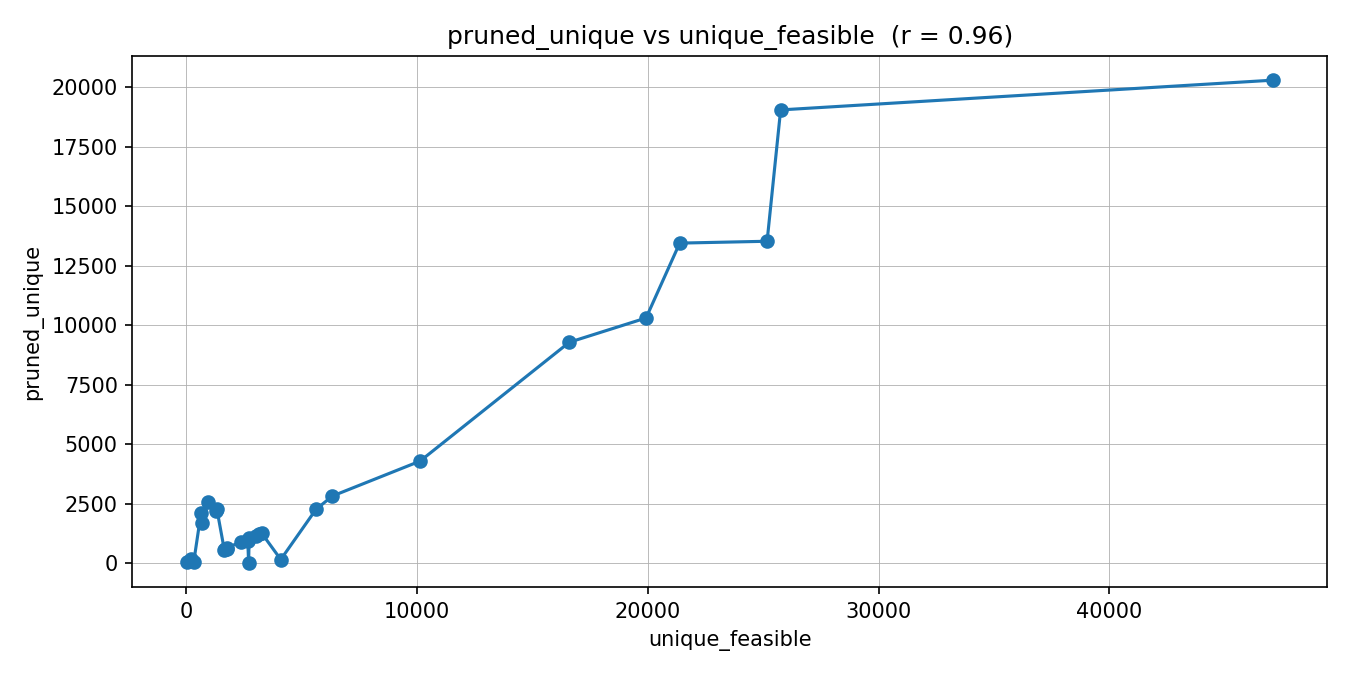}
\\ (a)
\end{minipage}
\begin{minipage}{0.48\columnwidth}
\centering
\includegraphics[width=\linewidth]{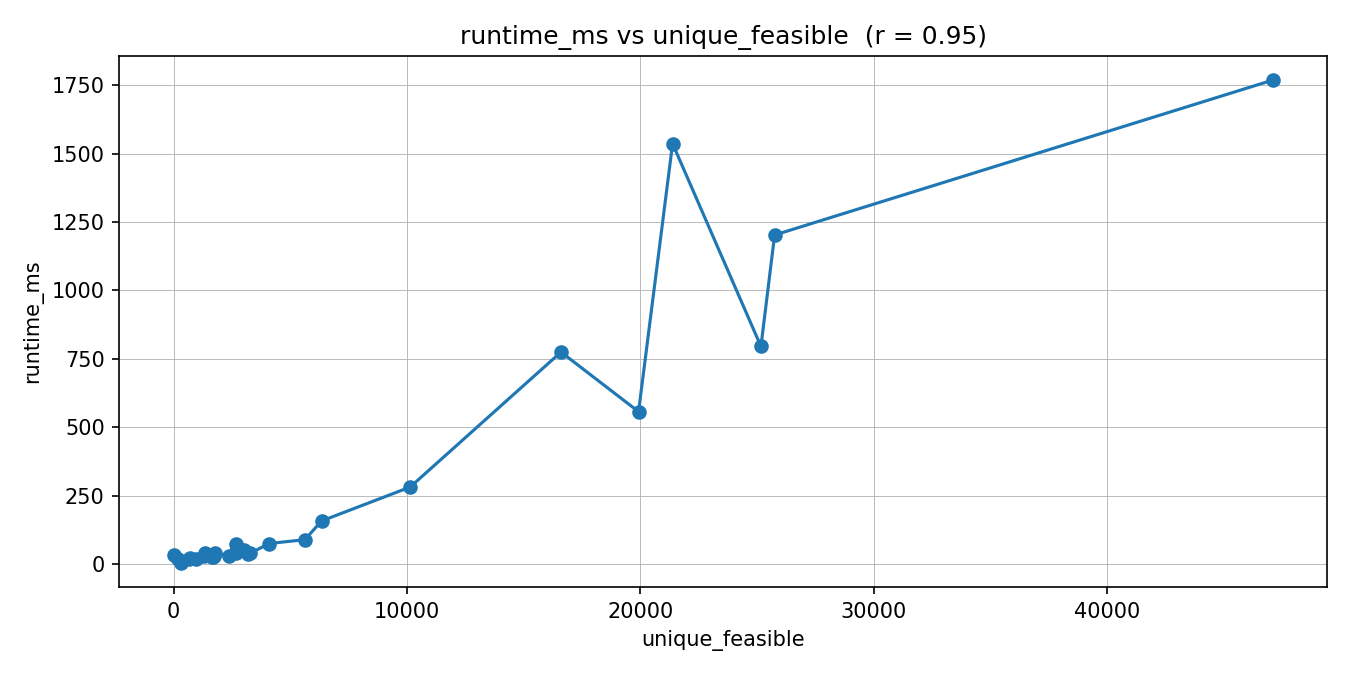}
\\ (b)
\end{minipage}

\vspace{1pt}

\begin{minipage}{0.48\columnwidth}
\centering
\includegraphics[width=\linewidth]{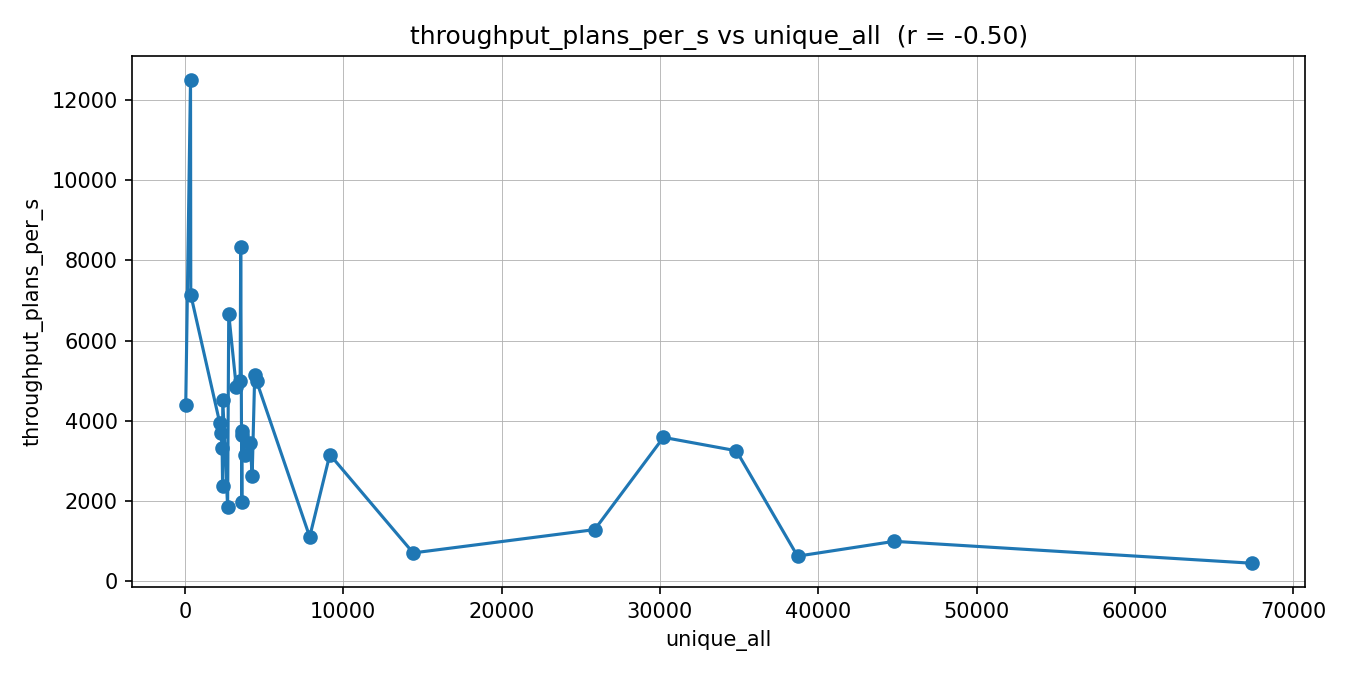}
\\ (c)
\end{minipage}
\begin{minipage}{0.48\columnwidth}
\centering
\includegraphics[width=\linewidth]{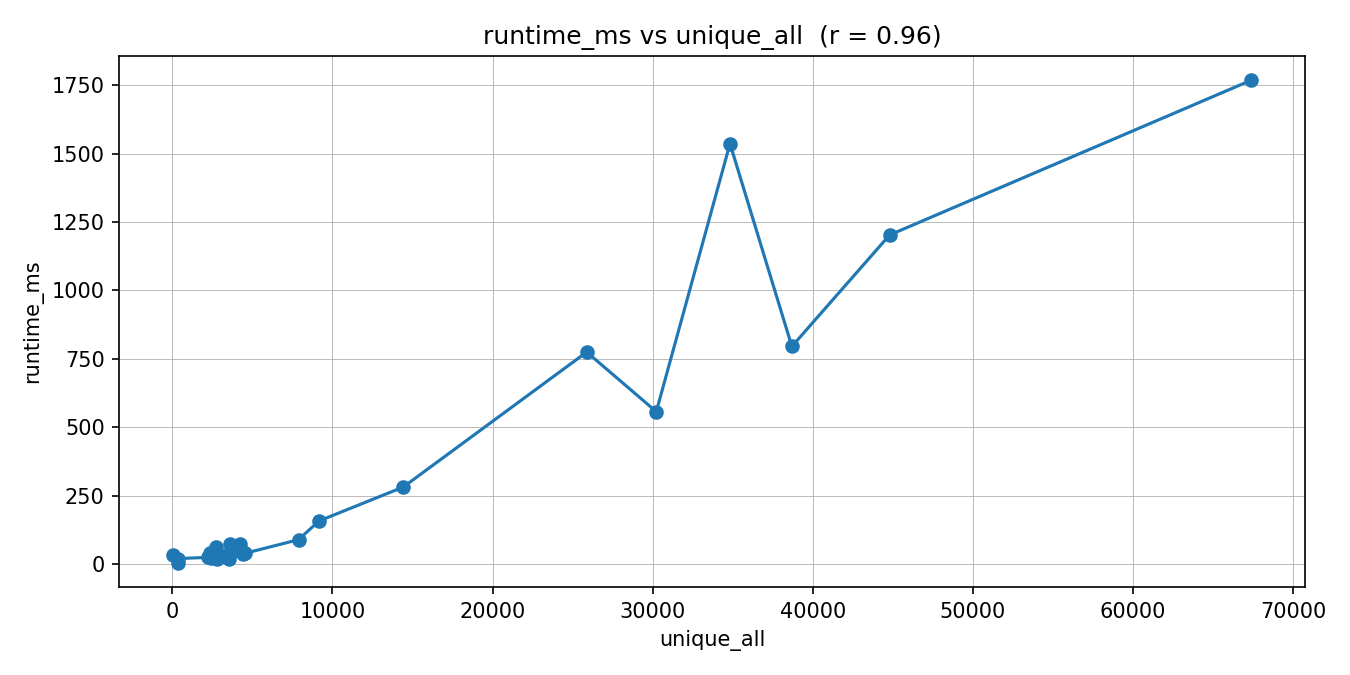}
\\ (d)
\end{minipage}

\vspace{1pt}

\begin{minipage}{0.48\columnwidth}
\centering
\includegraphics[width=\linewidth]{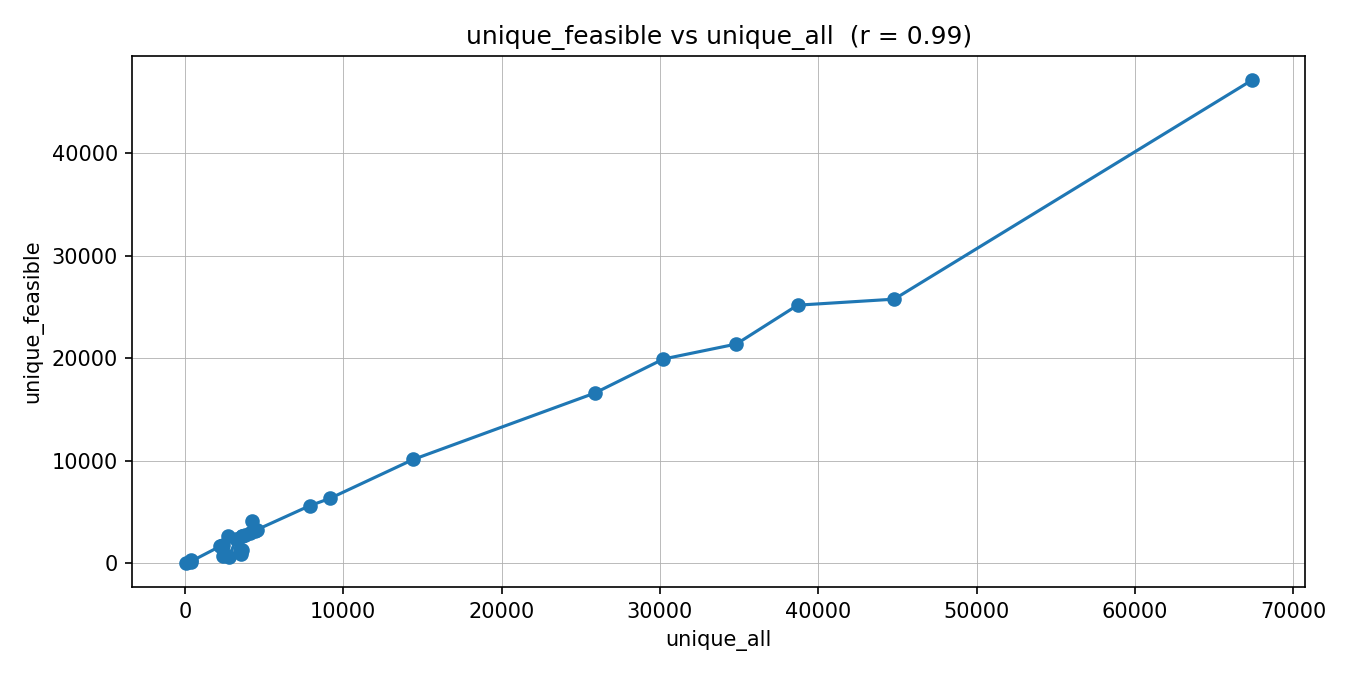}
\\ (e)
\end{minipage}
\begin{minipage}{0.48\columnwidth}
\centering
\includegraphics[width=\linewidth]{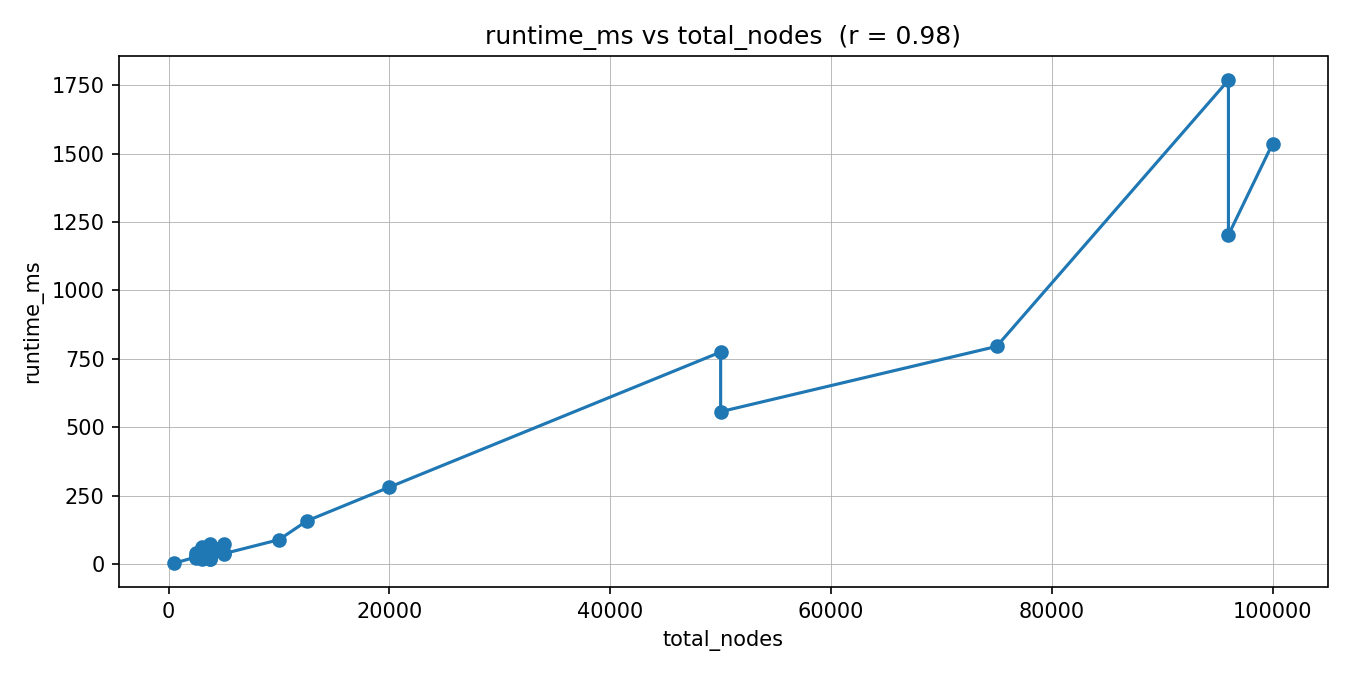}
\\ (f)
\end{minipage}

\caption{\scriptsize Performance metrics across multiple parameters. Each subplot shows pairwise correlation ($r$) between key variables. (a) pruned vs.\ feasible nodes, 
(b) runtime vs.\ feasible nodes, 
(c) throughput vs.\ total unique nodes, 
(d) runtime vs.\ total unique nodes, 
(e) feasible vs.\ total unique nodes, 
(f) runtime vs.\ total nodes.}
\label{fig:ppf_runtime_1}
\end{figure}

\section{Experimental Evaluation}
\label{sec:experiment}
We evaluated PPF construction (Algorithm~1) on six synthetic scenarios varying in structural overlap ($\rho_s$), feasibility ratio ($\rho_f$), and engine diversity ($\rho_e$), measuring redundancy reduction, pruning efficiency, and witness correctness. In Table~\ref{tab:ppf-eval}  Each scenario models a specific operational condition under which the algorithm is expected to perform differently. Together, they capture realistic workloads encountered during federated query optimization and planning for heterogeneous systems. The overall goal is to measure how efficiently the algorithm can (a) reduce redundancy through node packing, (b) prune infeasible branches early, and (c) maintain correctness through witness tracking. Below we summarize the six scenarios that were used to generate the workloads for the experiments presented in Table~\ref{tab:ppf-eval}.\\[4pt]
\textbf{Experimental Methodology and Analysis: } Fig.~\ref{fig:ppf_runtime_1}  confirms near-linear scaling: \textit{unique\_feasible} and \textit{pruned\_unique} strongly predict runtime ($|r|{>}0.7$); \textit{packed\_ratio} improves with diversity before plateauing. Together, \textit{unique\_feasible}, \textit{unique\_all}, and \textit{total\_nodes} guide pruning and scheduling. PPF compresses candidate plans by $4.2$--$11.7\times$; feasibility pruning removes $21$--$63\%$ of alternatives in $7.1$--$24.6$\,ms per query, confirming its role as a practical front-end for NL-driven heterogeneous query processing.\\[4pt]
\textbf{Baseline Comparison : } To contextualize the benefits of the PPF, we compare against two baselines that represent common approaches to handling multiple candidate plans in multimodel systems.\\[4pt]
\textbf{Baseline A: Naïve Enumeration.} The naïve strategy constructs each ILP independently and evaluates feasibility only at execution time. Given $n$ predicates with $k$ alternatives per predicate, this produces $O(k^n)$ candidate ILPs. In Example~3, eight ILPs contain more than 40 total operators, whereas the PPF contains only nine distinct nodes. Across the NL workload in Table~\ref{tab:baseline-compare}, naïve enumeration produced between 150 and 870 total operator instances per query, whereas the PPF reduced this to 5--12 nodes.\\[4pt]
\textbf{Baseline B: Memoization Without Annotations.}
A second baseline is a Volcano-style memo structure that merges operators based solely on their symbol and schema-level signature. This approach incorrectly merges nodes that are semantically incompatible (e.g., vector similarity computed over different embedding domains), leading to false feasible plans and delayed runtime failures. To quantify this effect, we disabled annotation-aware merging in our PPF implementation. 
Table~\ref{tab:baseline-compare} reports the number of ``incorrect merges'' (i.e., merges undone later by infeasibility checks). Between 28\% and 46\% of merged nodes in the annotation-agnostic memo were invalid, demonstrating that annotation-sensitive feasibility reasoning is necessary for correctness. 
\begin{table}[t]
\centering
\caption{Comparison to baseline memoization without annotations.}
\label{tab:baseline-compare}
\begin{tabular}{|c|c|c|}
\hline
\textbf{Query} & \textbf{Memo merges} & \textbf{Incorrect merges (\%)} \\
\hline
Q1 & 18 & 33\% \\
Q2 & 21 & 38\% \\
Q3 & 26 & 46\% \\
Q4 & 11 & 28\% \\
Q5 & 23 & 41\% \\
\hline
\end{tabular}
\end{table}
Together, these comparisons show that annotation-aware feasibility is crucial: PPF cuts plan sizes by up to $11\times$, while memo structures without annotations yield $28$--$46\%$ incorrect merges.
\section{Discussion and Conclusion:} 
We presented the Packed Plan Forest (PPF), a polynomially bounded structure that compactly encodes all feasible ILPs across heterogeneous query engines, supporting efficient feasibility checking via a bottom-up labeling algorithm. Experiments confirm polynomial scaling and exponential compression ratios. The PPF acts as a feasibility filter that precedes traditional rule-based or cost-based optimization. Once feasibility labeling completes, the remaining annotation configurations form a constraint-consistent search space that can be passed directly into a polystore cost-based optimizer (which is outside our scope). Feasibility certificates can further guide optimization by identifying operators that are infeasible for certain engines, suggesting early placement-based pruning. 

\bibliographystyle{splncs04}
\bibliography{mybib}

@article{LiJagadish2014VLDB,
  author    = {Fei Li and H. V. Jagadish},
  title     = {Constructing an Interactive Natural Language Interface for Relational Databases},
  journal   = {Proceedings of the VLDB Endowment},
  volume    = {8},
  number    = {1},
  pages     = {73--84},
  year      = {2014},
  publisher = {VLDB Endowment},
  doi       = {10.14778/2735461.2735468}
}

@article{Affolter2019VLDBJ,
  author    = {Katharina Affolter and Kurt Stockinger and Abraham Bernstein},
  title     = {A Survey of Natural Language Interfaces to Databases},
  journal   = {VLDB Journal},
  volume    = {28},
  number    = {5},
  pages     = {709--751},
  year      = {2019},
  publisher = {Springer},
  doi       = {10.1007/s00778-019-00557-9}
}

@inproceedings{Yaghmazadeh2017POPL,
  author    = {Navid Yaghmazadeh and Yuepeng Wang and Isil Dillig and Thomas Dillig},
  title     = {{SQLizer}: Query Synthesis from Natural Language},
  booktitle = {Proceedings of the 44th ACM SIGPLAN Symposium on Principles of Programming Languages (POPL)},
  pages     = {63--76},
  year      = {2017},
  publisher = {ACM},
  doi       = {10.1145/3009837.3009879}
}

@inproceedings{Guo2019ACL,
  author    = {Jiaqi Guo and Zecheng Zhan and Yan Gao and Yan Xiao and Jian-Guang Lou and Ting Liu and Dongmei Zhang},
  title     = {Towards Complex Text-to-{SQL} in Cross-Domain Databases},
  booktitle = {Proceedings of the 57th Annual Meeting of the Association for Computational Linguistics (ACL)},
  pages     = {4524--4535},
  year      = {2019},
  publisher = {ACL},
  doi       = {10.18653/v1/P19-1443},
  url       = {https://aclanthology.org/P19-1443}
}

@article{levy1996answering,
  title={Answering queries using views: A survey},
  author={Levy, Alon Y and Rajaraman, Anand and Ordille, Joann J},
  journal={VLDB Journal},
  volume={10},
  number={4},
  pages={270--294},
  year={1996}
}

@article{halevy2001answering,
  title={Answering queries using views: A survey},
  author={Halevy, Alon Y},
  journal={VLDB Journal},
  volume={10},
  number={4},
  pages={270--294},
  year={2001}
}

@inproceedings{benedikt2017query,
  title={Querying with access patterns and integrity constraints},
  author={Benedikt, Michael and Konstantinou, Nick and Ley-Wild, R and Murlak, Filip and Vrgoc, D},
  booktitle={Proceedings of the 20th International Conference on Extending Database Technology (EDBT)},
  pages={231--242},
  year={2017}
}

@article{stonebraker2018bigdawg,
  title={BigDAWG: A polystore system for analytics on heterogeneous data},
  author={Stonebraker, Michael and Balazinska, Magdalena and Cetintemel, Ugur and Cherniack, Mitch and Zdonik, Stanley},
  journal={Proceedings of the VLDB Endowment},
  volume={11},
  number={7},
  pages={819--831},
  year={2018}
}

@article{graefe1993volcano,
  title={The Volcano optimizer generator: Extensibility and efficient search},
  author={Graefe, Goetz and McKenna, William J},
  journal={Proceedings of ICDE},
  pages={209--218},
  year={1993}
}

@inproceedings{li2014constructing,
  title={Constructing natural language interfaces to databases},
  author={Li, Fei and Jagadish, HV},
  booktitle={Proceedings of the VLDB Endowment},
  volume={8},
  number={12},
  pages={73--84},
  year={2014}
}

@inproceedings{zhong2017seq2sql,
  title={Seq2SQL: Generating structured queries from natural language using reinforcement learning},
  author={Zhong, Victor and Xiong, Caiming and Socher, Richard},
  booktitle={Proceedings of ACL},
  year={2017}
}

@inproceedings{yaghmazadeh2017sqlizer,
  title={SQLizer: query synthesis from natural language},
  author={Yaghmazadeh, Navid and Wang, Yuepeng and Dillig, Isil and Dillig, Thomas},
  booktitle={Proceedings of POPL},
  pages={63--79},
  year={2017}
}

@inproceedings{yuan2023effective,
  title={An effective framework for enhancing query answering in a heterogeneous data lake},
  author={Yuan, Qin and Yuan, Ye and Wen, Zhenyu and Wang, He and Tang, Shiyuan},
  booktitle={Proceedings of the 46th International ACM SIGIR Conference on Research and Development in Information Retrieval},
  pages={770--780},
  year={2023}
}

@inproceedings{barret2023user,
  title={User-friendly exploration of highly heterogeneous data lakes},
  author={Barret, Nelly and Ebel, Simon and Galizzi, Th{\'e}o and Manolescu, Ioana and Mohanty, Madhulika},
  booktitle={International Conference on Cooperative Information Systems},
  pages={488--496},
  year={2023},
  organization={Springer}
}

@article{dar2019frameworks,
  title={Frameworks for querying databases using natural language: a literature review},
  author={Dar, Hafsa Shareef and Lali, M Ikramullah and Din, Moin Ul and Malik, Khalid Mahmood and Bukhari, Syed Ahmad Chan},
  journal={arXiv preprint arXiv:1909.01822},
  year={2019}
}

@inproceedings{dasgupta2016analytics,
  title={Analytics-driven data ingestion and derivation in the AWESOME polystore},
  author={Dasgupta, Subhasis and Coakley, Kevin and Gupta, Amarnath},
  booktitle={2016 IEEE International Conference on Big Data (Big Data)},
  pages={2555--2564},
  year={2016},
  organization={IEEE}
}

@inproceedings{kolev2016cloudmdsql,
  title={The cloudmdsql multistore system},
  author={Kolev, Boyan and Bondiombouy, Carlyna and Valduriez, Patrick and Jim{\'e}nez-Peris, Ricardo and Pau, Raquel and Pereira, Jos{\'e}},
  booktitle={Proceedings of the 2016 International Conference on Management of Data},
  pages={2113--2116},
  year={2016}
}

\end{document}